\documentclass[twocolumn,showpacs,preprintnumbers,amsmath,amssymb]{revtex4-1}

\usepackage{graphicx}
\usepackage{dcolumn}
\usepackage{bm}
\usepackage{amsfonts}
\usepackage{CJK}

\def\bequ{\begin{equation}}
\def\eequ{\end{equation}}
\def\be{\begin{equation}}
\def\ee{\end{equation}}

\begin{document}


\title{Maxwell perturbations on asymptotically anti-de Sitter spacetimes: \\
Generic boundary conditions and a new branch of quasinormal modes}

\author{Mengjie Wang}
\email{mengjie.wang@ua.pt}
\author{Carlos Herdeiro}
\email{herdeiro@ua.pt}
\author{Marco O. P. Sampaio}
\email{msampaio@ua.pt}
\affiliation{\vspace{2mm}Departamento de F\'\i sica da Universidade de Aveiro and CIDMA \\
Campus de Santiago, 3810-183 Aveiro, Portugal \vspace{1mm}}%

\date{\today}

\begin{abstract}
Perturbations of asymptotically Anti-de-Sitter ($AdS$) spacetimes are often considered by imposing \textit{field vanishing} boundary conditions (BCs) at the $AdS$ boundary. Such BCs, of Dirichlet-type, imply a vanishing energy flux at the boundary, but the converse is, generically, not true. Regarding $AdS$ as a gravitational box, we consider \textit{vanishing energy flux} (VEF) BCs as a more fundamental physical requirement and we show that these BCs can lead to a new branch of modes. As a concrete example, we consider \textit{Maxwell} perturbations on Kerr-$AdS$ black holes in the Teukolsky formalism, but our formulation applies also for other spin fields. Imposing VEF BCs, we find a set of \textit{two} Robin BCs, even for Schwarzschild-$AdS$ black holes.  The Robin BCs on the Teukolsky variables can be used to study quasinormal modes, superradiant instabilities and vector clouds. As a first application, we consider here the quasinormal modes of Schwarzschild-$AdS$ black holes. We find that one of the Robin BCs yields the quasinormal spectrum reported in the literature, while the other one unveils a new branch for the quasinormal spectrum.

\end{abstract}

\maketitle


\section{Introduction}
The stability of a black hole (BH) is a crucial question in BH physics. The reason is twofold. From the viewpoint of astrophysics, sufficiently stable BHs provide the best candidates to explain observations; $e.g$, high energy astrophysical processes like Active Galactic Nuclei (AGN). From the theoretical viewpoint, BH stability relates to the uniqueness theorems~\cite{Chrusciel:2012jk}, since BH instabilities may lead to new BH solutions when a \textit{zero-mode} of the instability exists; examples include new string solutions~\cite{Gubser:2001ac} due to the Gregory-Laflamme instability~\cite{Gregory:1993vy} and new asymptotically flat hairy rotating BH solutions~\cite{Herdeiro:2014goa,Herdeiro:2014ima,Herdeiro:2015gia} due to the superradiant instability~\cite{Press:1972zz}.

The problem of BH stability is typically tackled by introducing linear perturbations of test fields on a fixed background and studying either quasinormal modes or quasi-bound states ($cf.$ reviews~\cite{Berti:2009kk,Konoplya:2011qq,Cardoso:2014uka}). Quite remarkably, it has been shown that the equations of motion for linear perturbations of massless spin fields on four dimensional Kerr BHs both separate and decouple, yielding the cellebrated Teukolsky equation~\cite{Teukolsky:1973ha}. Subsequently, this equation has been generalized to rotating BHs with a cosmological constant in different contexts~\cite{Khanal:1983vb,Dias:2012pp,Wu:2003qc}.

To solve the Teukolsky equation, one has to assign physically relevant boundary conditions (BCs) which depend on the specific problem. In the context of quasinormal modes in asymptotically Anti-de Sitter ($AdS$) BHs, the most studied perturbations are those of scalar fields, for which \textit{field vanishing} BCs are usually imposed, see $e.g.$~\cite{Uchikata:2009zz}. For other spin fields, the problem has only been partly addressed. The quasinormal modes for the Maxwell field and gravitational field in Schwarzschild-$AdS$ BHs have been obtained using the Regge-Wheeler method~\cite{Regge:1957td}, instead of the Teukolsky equation, in~\cite{Cardoso:2001bb,Cardoso:2003cj}, exploring the spherical symmetry of the background. Additionally, these works impose field vanishing BCs. For non-spherically symmetric backgrounds, like in Kerr-$AdS$ BHs, one must, however, use the Teukolsky formalism  and, since this formalism uses a different set of variables, it is not obvious how to impose BCs for non-zero spin fields. Recently, superradiant instabilities of the gravitational field on Kerr-$AdS$ BHs have been studied~\cite{Cardoso:2013pza} with BCs~\cite{Dias:2013sdc} chosen as to preserve the asymptotic global $AdS$ structure of the background\footnote{Such BCs, as argued in~\cite{Holzegel:2015swa}, do not appear to give rise to a well posed time evolution.}. Furthermore, it was proved in~\cite{Cardoso:2013pza} that the BCs in~\cite{Dias:2013sdc} yield vanishing energy flux at the asymptotic boundary.

The $AdS$ boundary may be regarded as a perfectly reflecting mirror, in the sense that no energy flux can cross the asymptotic boundary. We will take this viewpoint as our basic principle for imposing BCs for linear perturbations of asymptotically $AdS$ spacetimes. It suggests taking \textit{vanishing energy flux} (VEF) BCs, which should be contrasted to the \textit{field vanishing} BCs we mentioned before. In this paper we will illustrate how these BCs, based on this simple physical principle, can lead to new results, using the Maxwell field as an example.

We present a framework to calculate VEF BCs for the Maxwell field on Kerr-$AdS$ BHs, when using the Teukolsky equation. From the VEF BCs we get a set of \textit{two} Robin BCs. These are determined by a linear combination of the Teukolsky variables and their derivatives, for the Maxwell field in Kerr-$AdS$ BHs. Interestingly, these two conditions are different even in Schwarzschild-$AdS$ BHs; thus, in this paper, we will focus on the latter background. We then observe that one of the Robin BCs recovers the results shown in~\cite{Cardoso:2001bb,Cardoso:2003cj,Berti:2003ud}, but the other one leads to a hitherto unknown branch of quasinormal modes. Other applications of the VEF BCs, such as superradiant instabilities and vector clouds of Kerr-$AdS$ BHs will be reported elsewhere~\cite{wangherdeiro}.

The formulation we present to construct BCs applies to both the Maxwell field as well as for other spin fields. For a scalar field, however, the VEF BCs yield equivalent results to the Dirichlet BCs typically used. It would be interesting to consider the VEF BCs for other spin fields as well, especially for a gravitational field, to inquire if they are \textit{equivalent} to the BCs in~\cite{Dias:2013sdc}.

The structure of this paper is organized as follows. In Section~\ref{seceq} we introduce the Kerr-$AdS$ background geometry, and the Teukolsky equation for the Maxwell field which will be studied in this paper. In Section~\ref{secbc} we show how to get the VEF BCs in the Teukolsky formalism. In Section~\ref{secSch} we apply these BCs to Schwarzschild-$AdS$ BHs, and discuss the two branches of quasinormal frequencies, one of them already reported in the literature, and a new one which has not been explored yet. Final remarks and conclusions are presented in the last section.

\section{background geometry and the field equation}
\label{seceq}
For a self-contained presentation, in this section we briefly review some basic properties of Kerr-$AdS$ BHs, and the Teukolsky equation for the Maxwell field on this background geometry.
We consider the following line element for a Kerr-$AdS$ BH
\begin{eqnarray}
ds^2=&-&\rho^2\left(\dfrac{dr^2}{\Delta_r}+\dfrac{d\theta^2}{\Delta_\theta}\right)+\dfrac{\Delta_r}{\rho^2\Xi^2}\Big(dt-a\sin^2\theta d\varphi\Big)^2\nonumber\\
&-&\dfrac{\Delta_\theta \sin^2\theta}{\rho^2\Xi^2}\Big(adt-(r^2+a^2)d\varphi\Big)^2 \;,\label{metric}
\end{eqnarray}
with metric functions
\begin{eqnarray}
\rho^2&=&r^2+a^2\cos^2\theta\;,\;\;\Delta_r=(r^2+a^2)\left(1+\frac{r^2}{L^2}\right)-2Mr\;,\nonumber\\
\Delta_\theta&=&1-\dfrac{a^2\cos^2\theta}{L^2}\;,\;\;\Xi=1-\dfrac{a^2}{L^2}\;,\label{metricfunc}
\end{eqnarray}
where $L$ is the $AdS$ radius and parameters $M$, $a$ are related to the BH energy $E$ and angular momentum $J$.
In this frame, the angular velocity of the event horizon and the Hawking temperature are given by
\begin{eqnarray}
\Omega_H&=&\dfrac{a}{r_+^2+a^2}\;,\label{Horvel}\\
T_H&=&\dfrac{1}{\Xi}\left[\dfrac{r_+}{2\pi}\left(1+\dfrac{r_+^2}{L^2}\right)\dfrac{1}{r_+^2+a^2}-\dfrac{1}{4\pi r_+}\left(1-\dfrac{r_+^2}{L^2}\right)\right]\;,\nonumber\\\label{Temp}
\end{eqnarray}
where the event horizon $r_+$ is determined as the largest root of $\Delta_r(r_+)=0$. For a given $r_+$, the mass parameter $M$ can be expressed as
\begin{equation}
M=\dfrac{(r_+^2+a^2)(L^2+r_+^2)}{2r_+L^2}\;.\nonumber
\end{equation}
For non-extremal BHs and to avoid singularities, we shall constrain the rotation parameter $a$~\cite{Cardoso:2013pza}
\begin{eqnarray}
&&\dfrac{a}{L}\leq \dfrac{r_+}{L} \sqrt{\dfrac{3r_+^2+L^2}{L^2-r_+^2}}\;,\;\;\;\;\;\text{for}\;\; \dfrac{r_+}{L}<\dfrac{1}{\sqrt{3}}\;,\nonumber\\
&&\dfrac{a}{L}<1\;,\;\;\;\;\;\;\;\;\;\;\;\;\;\;\;\;\;\;\;\;\;\;\;\;\;\;\;\text{for}\;\; \dfrac{r_+}{L}\geq\dfrac{1}{\sqrt{3}}\;.\nonumber
\end{eqnarray}
\\
A linear perturbation equation for massless spin fields on a Kerr BH was worked out by Teukolsky in a pioneer work~\cite{Teukolsky:1973ha}, and was generalized to a Kerr-$dS$ BH later~\cite{Khanal:1983vb}. Recently the analogous equation was derived for a Kerr-$AdS$ BH~\cite{Dias:2012pp}. In the following we outline the equations for the radial and angular parts of the master field describing a spin $s$ perturbation, without a detailed derivation. For the case of interest herein, the spin parameter is $s=\pm1$.
\\
The radial equation is
\begin{equation}
\Delta_r^{-s}\dfrac{d}{dr}\left(\Delta_r^{s+1}\dfrac{d R_{s}(r)}{dr}\right)+H(r)R_{s}(r)=0\;,\label{radialeq}
\end{equation}
with
\begin{eqnarray}
H(r)=\dfrac{K_r^2-i s K_r \Delta_r^\prime}{\Delta_r}+2isK_r^\prime+\dfrac{s+|s|}{2}\Delta_r^{\prime\prime}
+\dfrac{a^2}{L^2}-\lambda\;,\nonumber
\end{eqnarray}
where
\begin{eqnarray}
K_r=[\omega(r^2+a^2)-am]\Xi\;.\label{Kreq}
\end{eqnarray}
The angular equation is
\begin{equation}
\dfrac{d}{du}\left(\Delta_u\dfrac{dS_{lm}}{du}\right)+A(u)S_{lm}=0\;,\label{angulareq}
\end{equation}
with $u=\cos\theta$, and
\begin{equation}
A(u)=-\dfrac{K_u^2}{\Delta_u}-4smu\dfrac{\Xi}{1-u^2}+\lambda-|s|-2(1-u^2)\dfrac{a^2}{L^2}\;,\nonumber
\end{equation}
where
\begin{eqnarray}
K_u&=&\Big(\omega a (1-u^2)+(s u-m)\Big)\Xi\;,\nonumber\\
\Delta_u&=&(1-u^2)\left(1-\dfrac{a^2}{L^2}u^2\right)\;.\nonumber
\end{eqnarray}

\section{boundary conditions}
\label{secbc}
To solve a differential equation, like the radial equation~\eqref{radialeq} and the angular equation~\eqref{angulareq}, one has to impose physically relevant BCs. For the angular equation~\eqref{angulareq}, one usually requires its solutions to be regular at the singular points $\theta=0$ and $\theta=\pi$. This determines uniquely the set of angular functions labelled by $\ell$ and $m$. For the radial equation~\eqref{radialeq}, we have to impose conditions both at the horizon and at infinity. At the horizon, ingoing BCs are imposed. At infinity, however, the BCs are more subtle. For the often studied case of a scalar field on Kerr-$AdS$ BHs,  the BCs typically imposed require the field itself to vanish~\cite{Uchikata:2009zz,Cardoso:2003cj}, when looking for quasinormal modes. For the Maxwell field and in the Teukolsky formalism, the asymptotic BCs have not been explored yet. In this section, we are going to discuss them for the general Kerr-$AdS$ background.

We propose that in the Teukolsky formalism, when looking for quasinormal modes of the Maxwell field on Kerr-$AdS$ BHs, VEF BCs should be imposed, following the spirit that the $AdS$ boundary is a perfectly reflecting mirror so that no energy flux can cross it. For the particular case of the electromagnetic field, these BCs create an analogy between the $AdS$ boundary and a perfect conductor. Actually the conductor condition for the Maxwell field has been considered in Kerr-mirror system~\cite{Brito:2015oca}. But the VEF BCs, which for a scalar field can yield both standard Dirichlet and Neumann BCs and for a Maxwell field can yield perfectly conducting BCs, are a  general principle for any spin field, based on a sound physical rationale.

\bigskip

The energy-momentum tensor for the Maxwell field is
\begin{equation}
T_{\mu \nu}=F_{\mu\sigma}F^\sigma_{\;\;\;\nu}+\dfrac{1}{4}g_{\mu\nu}F^2\;,\label{EMTensor}
\end{equation}
with the Maxwell tensor $F_{\mu\nu}$~\cite{Teukolsky:1973ha}
\begin{eqnarray}
F_{\mu\nu}=&&2\left(\phi_1(n_{[\mu} l_{\nu]}+m_{[\mu} m^\ast_{\nu]})+\phi_2 l_{[\mu} m_{\nu]}+\phi_0 m^\ast_{[\mu} n_{\nu]}\right)\nonumber\\
&&+c.c\;,\nonumber
\end{eqnarray}
where square brackets on subscripts stand for antisymmetrization, and $c.c$ stands for complex conjugate of the preceding terms. The tetrad is constructed from the line element in Eq.~\eqref{metric}, with definition
\begin{eqnarray}
l^\mu&=&\left(\dfrac{(r^2+a^2)\Xi}{\Delta_r},1,0,\dfrac{a\Xi}{\Delta_r}\right)\;,\nonumber\\
n^\mu&=&\dfrac{1}{2\rho^2}\Big((r^2+a^2)\Xi,-\Delta_r,0,a\Xi\Big)\;,\nonumber\\
m^\mu&=&\dfrac{1}{\sqrt{2\Delta_\theta}\bar{\rho}}\left(ia\Xi\sin\theta,0,\Delta_\theta,\dfrac{i\Xi}{\sin\theta}\right)\;,\nonumber
\end{eqnarray}
where $\bar{\rho}=r+ia\cos\theta$. \\
The Maxwell scalars are defined as
\begin{eqnarray}
&&\phi_0=F_{\mu\nu}l^\mu m^\nu\;,\;\;\;\phi_1=\dfrac{1}{2}F_{\mu\nu}(l^\mu n^\nu+m^{\ast\mu}m^\nu)\;,\nonumber\\
&&\phi_2=F_{\mu\nu}m^{\ast\mu}n^\nu\;,\nonumber
\end{eqnarray}
where $m^{\ast\mu}=(m^\mu)^\ast$.\\
We are now able to calculate the radial energy flux $T^r_{\;\;t}$, by substituting all of the above ingredients into Eq.~\eqref{EMTensor}, which gives
\begin{equation}
T^r_{\;\;t}=T^r_{\;\;t,\;\uppercase\expandafter{\romannumeral1}}+T^r_{\;\;t,\;\uppercase\expandafter{\romannumeral2}}\;,\nonumber
\end{equation}
where
\begin{equation}
T^r_{\;\;t,\;\uppercase\expandafter{\romannumeral1}}=\dfrac{1}{2\Xi}\left(4|\phi_2|^2-\dfrac{\Delta_r^2}{\rho^4}|\phi_0|^2\right)\;,\label{rtcom}
\end{equation}
while $T^r_{\;\;t,\;\uppercase\expandafter{\romannumeral2}}$ becomes irrelevant at infinity, so we do not show its expression here.\\
Then we decompose the Maxwell scalars as
\begin{eqnarray}
\phi_0&=&e^{-i\omega t+im\varphi} R_{+1}(r)S_{+1}(\theta)\;,\nonumber\\
\phi_2&=&\dfrac{B}{2(\bar{\rho}^\ast)^2}e^{-i\omega t+im\varphi} R_{-1}(r)S_{-1}(\theta)\;,\label{fielddeccom}
\end{eqnarray}
where $B$ is a positive root of~\cite{Wu:2003qc}
\begin{equation}
B^2=\lambda^2-4\Xi^2\omega(\omega a^2-ma)\;,\nonumber
\end{equation}
such that the Starobinsky-Teukolsky identities are satisfied~\cite{Wu:2003qc}
\begin{eqnarray}
R_{+1}&=&\left(\dfrac{d}{dr}-\dfrac{iK_r}{\Delta_r}\right)\left(\dfrac{d}{dr}-\dfrac{iK_r}{\Delta_r}\right)R_{-1}\;,\label{identity1}\\
B^2R_{-1}&=&\Delta_r\left(\dfrac{d}{dr}+\dfrac{iK_r}{\Delta_r}\right)\left(\dfrac{d}{dr}+\dfrac{iK_r}{\Delta_r}\right)P_{+1}\;,\label{identity2}
\end{eqnarray}
where $K_r$ is given by Eq.~\eqref{Kreq}, $P_{+1}=\Delta_r R_{+1}$, and $R_{\pm1}(r)$ and $S_{\pm1}(\theta)$ obey the radial equation~\eqref{radialeq} and the angular equation~\eqref{angulareq}, respectively.

With the fields decomposition in Eq.~\eqref{fielddeccom}, integrating $T^r_{\;\;t,\;\uppercase\expandafter{\romannumeral1}}$ over a sphere, we obtain the energy flux
\begin{eqnarray}
\mathcal{F}|_r&=&\int_{S^2} \sin\theta d\theta d\varphi\; r^2 T^r_{\;\;t,\;\uppercase\expandafter{\romannumeral1}}\;\nonumber\\
&=&\dfrac{r^2}{2\;\Xi\;\rho^4}(B^2|R_{-1}|^2-\Delta_r^2|R_{+1}|^2)\;,\label{flux}
\end{eqnarray}
up to an irrelevant normalization, and the angular functions $S_{\pm1}(\theta)$ are normalized
\begin{equation}
\int_0^\pi\sin\theta d\theta\; |S_{\pm1}(\theta)|^2=1\;.\nonumber
\end{equation}
To get the asymptotic boundary condition for $R_{-1}$, we expand Eq.~\eqref{radialeq} with $s=-1$ asymptotically as
\begin{equation}
R_{-1} \sim \;\alpha^{-} r+\beta^{-}+\mathcal{O}(r^{-1})\;,\label{asysol}
\end{equation}
where $\alpha^{-}$ and $\beta^{-}$ are two integration constants.
Keeping in mind the Starobinsky-Teukolsky identities~\eqref{identity1}, making use of the radial equation~\eqref{radialeq} with $s=-1$ and the asymptotic expansion in Eq.~\eqref{asysol}, at the asymptotic boundary, the energy flux in Eq.~\eqref{flux} becomes
\begin{eqnarray}
\mathcal{F}|_{r,\infty}=B^2|\alpha^{-}|^2-|(\lambda-2\omega^2\Xi^2L^2)\alpha^{-}+2i\beta^{-}\omega\Xi|^2\;,\nonumber\\\label{fluxinf}
\end{eqnarray}
where an overall proportional constant has been ignored.
To impose the VEF BCs, $i.e.$ $\mathcal{F}|_{r,\infty}=0$, we have
\begin{equation}
B^2|\alpha^{-}|^2-|(\lambda-2\omega^2\Xi^2L^2)\alpha^{-}+2i\beta^{-}\omega\Xi|^2=0\;.
\end{equation}
Note that $\alpha^{-}$ and $\beta^{-}$ are two independent integration constants; we can rescale them so that the modulus in the above equation can be dropped~\footnote{Actually there still might be a phase factor between these two constants, but this phase factor can be fixed through calculating the normal modes.}. Then it is easy to solve this quadratic equation obtaining the two solutions
\begin{equation}
\dfrac{\alpha^{-}}{\beta^{-}}=\dfrac{2i\omega\Xi}{\pm B-\lambda+2\omega^2\Xi^2L^2}\;.\label{bc}
\end{equation}
We have also checked that, the angular momentum flux of the Maxwell field vanishes asymptotically if the above boundary conditions are satisfied, similarly to the gravitational case~\cite{Cardoso:2013pza}.

For Schwarzschild-$AdS$ BHs, Eq.~\eqref{bc} simplifies to
\begin{equation}
\dfrac{\alpha^{-}}{\beta^{-}}=\dfrac{i}{\omega L^2}\;,\;\;\;\dfrac{\alpha^{-}}{\beta^{-}}=\dfrac{i\omega}{-\ell(\ell+1)+\omega^2L^2}\; .\label{bcSAdS}
\end{equation}
These are, apparently, two distinct Robin BCs, but at this moment it is unclear if they lead to physically different modes or if they are isospectral.

We can also follow the same procedure to calculate BCs for the Teukolsky equation with $s=+1$. Instead of using $R_{+1}(r)$, we use $P_{+1}(r)$ for convenience, which relates to $R_{+1}(r)$ through $P_{+1}(r)=\Delta_r R_{+1}(r)$. As before, we expand $P_{+1}(r)$ from Eq.~\eqref{radialeq} with $s=+1$ asymptotically
\begin{equation}
P_{+1} \sim \;\alpha^{+} r+\beta^{+}+\mathcal{O}(r^{-1})\;,\label{asysol2}
\end{equation}
where $\alpha^{+}$ and $\beta^{+}$ are two integration constants. Using the Starobinsky-Teukolsky identity in Eq.~\eqref{identity2}, the asymptotic expansion in Eq.~\eqref{asysol2}, the Teukolsky equation with $s=+1$ in Eq.~\eqref{radialeq} and the transformation $P_{+1}(r)=\Delta_r R_{+1}(r)$, then Eq.~\eqref{flux} gives the  conditions
\begin{equation}
\dfrac{\alpha^{+}}{\beta^{+}}=-\dfrac{2i\omega\Xi}{\pm B-\lambda+2\omega^2\Xi^2L^2}\;,\label{bc2}
\end{equation}
after imposing the VEF BCs. Comparing the two BCs in~\eqref{bc} and in~\eqref{bc2}, we find that there is only a sign difference, or in other words, they are complex conjugate to each other. This is the consequence that $P_{+1}(r)$ and $R_{-1}(r)$ are proportional to complex conjugate functions of each other. We have checked that solving the radial equation~\eqref{radialeq} for $s=-1$ and $s=+1$ with the corresponding BCs~\eqref{bc} and~\eqref{bc2}, for Schwarzschild-$AdS$ BHs, the same quasinormal frequencies are obtained, which is consistent with the argument that these two Teukolsky equations encode the same information. Thus, for concreteness and without loss of generality, in the following we specify $s=-1$, and consider the corresponding BCs.

\section{Maxwell quasinormal modes for Schwarzschild-$AdS$ BHs}
\label{secSch}
We shall now apply the VEF BCs to Maxwell perturbations on Schwarzschild-$AdS$ BHs, in the Teukolsky formalism. We show that even in this simpler case, there is a new branch of quasinormal modes which has not been explored yet. Before that, however, we calculate normal modes in pure $AdS$ spacetime, which not only illustrates how the BCs work, but also provides an initial guess for the later numerical calculations of quasinormal modes. In the pure $AdS$ case the spectra obtained from the two different Robin BCs are isospectral (up to one mode).

\subsection{Normal modes}
The normal frequencies for the Maxwell field on an empty $AdS$ background can be obtained analytically. The radial Teukolsky equation~\eqref{radialeq} can be simplified in this case to
\begin{eqnarray}
&&\Delta_rR_{-1}''(r)+\left(\dfrac{K_r^2+i K_r \Delta_r^\prime}{\Delta_r}-2iK_r^\prime
-\ell(\ell+1)\right)R_{-1}(r)\nonumber\\
&&=0\;,\label{fareq1}
\end{eqnarray}
with
\begin{equation}
\Delta_r= r^2 \left(1+\dfrac{r^2}{L^2}\right)\;,\;\;\;\;\;\;K_r=\omega r^2 .\nonumber
\end{equation}
The general solution for Eq.~\eqref{fareq1} is
\begin{eqnarray}
&&R_{-1}=\;r^{\ell+1}(r-iL)^{\frac{\omega L}{2}}(r+iL)^{-\ell-\frac{\omega L}{2}}\Big[C_1F\Big(\ell,\ell+1\Big.\Big.\nonumber \\ && \Big.\Big.+\omega L,2\ell+2;\dfrac{2r}{r+iL}\Big)+C_2(-1)^{2\ell+1}2^{-2\ell-1}\left(1+\dfrac{iL}{r}\right)^{2\ell+1}\Big.\nonumber \\ && \Big.F\Big(-\ell-1,-\ell+\omega L,-2\ell;\dfrac{2r}{r+iL}\Big)\Big]\;,\label{farsol}
\end{eqnarray}
where $F(a,b,c;z)$ is the hypergeometric function, $C_1$ and $C_2$ are two integration constants with dimension of inverse length. These are related to each other by the BCs through expanding Eq.~\eqref{farsol} at large $r$:
\begin{itemize}
\item[$\bullet$] Imposing the first of the two BCs in Eq.~\eqref{bcSAdS}, one gets a first relation between $C_1$ and $C_2$
\begin{equation}
\dfrac{C_2}{C_1}=-2^{2\ell+1}\dfrac{\ell}{\ell+1}\dfrac{F(\ell+1,\ell+1+\omega L,2\ell+2;2)}{F(-\ell,-\ell+\omega L,-2\ell;2)}\;.\label{c1c2bc1}
\end{equation}
\item[$\bullet$] Imposing the second, of the two BCs in Eq.~\eqref{bcSAdS}, on the other hand, one gets a second relation between $C_1$ and $C_2$
\begin{equation}
\dfrac{C_2}{C_1}=2^{2\ell+1}\left(\dfrac{\ell}{\ell+1}\right)^2\dfrac{\ell+1+\omega L}{\ell-\omega L}\dfrac{\mathcal{A}_1}{\mathcal{A}_2}\;,\label{c1c2bc2}
\end{equation}
where
\begin{eqnarray}
\mathcal{A}_1=&&(\ell+1)F(\ell,\ell+1+\omega L,2\ell+2;2)+\omega L F(\ell+1,\nonumber\\&&\ell+2+\omega L,2\ell+3;2)\;,\nonumber\\
\mathcal{A}_2=&&\ell F(-\ell-1,-\ell+\omega L,-2\ell;2)-\omega L F(-\ell,-\ell+1\nonumber\\&&+\omega L,1-2\ell;2)\;.
\end{eqnarray}
\end{itemize}
Then from the small $r$ behavior of Eq.~\eqref{farsol}
\begin{equation}
R_{-1}\;\sim\;\dfrac{-i L^{\ell+1} C_2}{r^\ell}+ (-1)^\ell2^{2\ell+1}L^{-\ell} C_1 r^{\ell+1}\;,\label{farsolnear}
\end{equation}
we have to set $C_2=0$ in order to get a regular solution at the origin. This regularity condition picks the normal modes, from Eqs.~\eqref{c1c2bc1} and~\eqref{c1c2bc2}:
\begin{eqnarray}
&&F(\ell+1,\ell+1+\omega L,2\ell+2;2)=0\;,\nonumber\\
&&\Rightarrow\;\;\omega_{1,N}L=2N+\ell+2\;,\label{normalmode1}\\
&&\mathcal{A}_1=0\;,\nonumber\\&&\Rightarrow\;\;\omega_{2,N}L=2N+\ell+1\;,\label{normalmode2}
\end{eqnarray}
where $N=0,1,2,\cdot\cdot\cdot$, and $\ell=1,2,3,\cdot\cdot\cdot$. Observe that, as announced before, the two sets of frequencies are isospectral, up to one mode. Observe also that these two normal modes are the same with the gravitational case, as shown in~\cite{Cardoso:2013pza}.

\subsection{Quasinormal modes}
When the BH effects are taken into account, we cannot solve the radial Teukolsky equation analytically as before. So we are now going to look for quasinormal modes of Schwarzschild-$AdS$ BHs by solving the Teukolsky equation numerically.

As we mentioned before, the quasinormal modes for the Maxwell field on Schwarzschild-$AdS$ BHs have been studied using the Regge-Wheeler formalism~\cite{Cardoso:2003cj}. Here we will tackle the same problem in the Teukolsky formalism, imposing the BCs discussed in the Section~\ref{secbc}. We find that:
\begin{itemize}
\item[$\bullet$] when the first of the two BCs in Eq.~\eqref{bcSAdS} is imposed, we recover the results given in the literature~\cite{Cardoso:2003cj,Berti:2003ud};
\item[$\bullet$] when the second of the two BCs in Eq.~\eqref{bcSAdS} is imposed, there is one new branch of quasinormal modes.
\end{itemize}
To be complete and comparative, we will show both results in the following. In the numerical calculations all physical quantities are normalized by the $AdS$ radius $L$ and we set $L=1$. Also, observe that we use $\omega_1$ ($\omega_2$) to represent the quasinormal frequency corresponding to the first (second) BCs.

\bigskip

To solve the radial equation~\eqref{radialeq} with $s=-1$, we use a direct integration method, adapted from~\cite{Herdeiro:2011uu,Wang:2012tk,Sampaio:2014swa}. Firstly, we use Frobenius' method to expand $R_{-1}$ close to the event horizon
\begin{equation}
R_{-1}=(r-r_+)^\rho \sum_{j=0}^\infty c_j\;(r-r_+)^j\;,\nonumber
\end{equation}
with the ingoing boundary condition
\begin{equation}
\rho=1-\dfrac{i\omega r_+}{1+3r_+^2}\;,\nonumber
\end{equation}
and initialize the integration of Eq.~\eqref{radialeq} therein. The series expansion coefficients $c_j$ can be derived directly after inserting these expansions into Eq.~\eqref{radialeq}.
At infinity, the asymptotic behavior of $R_{-1}$ has been given in Eq.~\eqref{asysol}, where two coefficients, $\alpha^{-}$ and $\beta^{-}$, can be extracted from $R_{-1}$ and its first derivative. For that purpose, we can define two new fields $\left\{\chi,\psi\right\}$, which will asymptote respectively to $\left\{\alpha^{-},\beta^{-}\right\}$, at infinity. Such a transformation can be written in a matrix form by defining the vector $\mathbf{\Psi}^T=(\chi,\psi)$ for the new fields, and another vector $\mathbf{V}^T=(R_{-1},\frac{d}{dr}R_{-1})$ for the original field and its derivative. Then the transformation is given in terms of an $r$-dependent matrix $\mathbf{T}$ defined through
\begin{equation}
\mathbf{V}= \left(\begin{array}{cc} r & 1 \vspace{2mm}\\ 1 & 0 \end{array}\right) \mathbf{\Psi} \equiv \mathbf{T} \mathbf{\Psi} \;.\nonumber
\end{equation}
To obtain a first order system of ODEs for the new fields, we first define a matrix $\mathbf{X}$ through
\begin{equation}
\dfrac{d\mathbf{V}}{dr}=\mathbf{X}\mathbf{V} \; ,
\end{equation}
where $\mathbf{X}$ can be read out from the original radial equation~\eqref{radialeq}.
Then we obtain
\begin{equation}
\dfrac{d\mathbf{\Psi}}{dr}=\mathbf{T}^{-1}\left(\mathbf{X}\mathbf{T}-\dfrac{d\mathbf{T}}{dr}\right) \mathbf{\Psi} \;,\label{radialmatrix}
\end{equation}
which is the equation we are going to solve numerically.
With this numerical procedure and the BCs given in Eq.~\eqref{bcSAdS}, we calculate quasinormal frequencies.

In Table~\ref{EM1}, we list a few fundamental $(N=0)$ quasinormal frequencies of $\omega_1$ (with $\ell=1$) and $\omega_2$ (with $\ell=2$), for different BH sizes. As mentioned above, the normal modes displayed in Eqs.~\eqref{normalmode1} and~\eqref{normalmode2}, are isospectral under the mapping
\begin{equation}
\ell_2\leftrightarrow \ell_1+1 \ ,
\end{equation}
except one mode for $\omega_2$, where $\ell_1$ and $\ell_2$ refer to the angular momentum quantum number in the spectrum of $\omega_1$ and $\omega_2$. The presence of a BH, however, breaks the isospectrality. To show this, we present in Table~\ref{EM1}, the two sets of quasinormal frequencies, with $\ell_1=1$ and $\ell_2=2$, respectively.
One observes that the degeneracy between $\omega_1$ and $\omega_2$ gets broken, especially in the small BH and intermediate BH regimes. For large BHs, these two modes are, again, almost isospectral, which seems to be a general feature for any type of perturbation~\cite{Berti:2003ud,Cardoso:2003cj}.
Furthermore, for large BHs, the real part of the frequency for either of the sets vanishes, while the imaginary part scales linearly with the BH size $r_+$. This scaling can be equally stated in terms of the Hawking temperature, which relates to the BH size through $T_H=3r_+/(4\pi L^2)$ for large BHs, supporting the arguments given in~\cite{Horowitz:1999jd}, where a similar linear relation was found for scalar fields.
We remark that the numerical data for $\omega_1$ displayed in Table~\ref{EM1} coincides with the numerical results presented in~\cite{Cardoso:2001bb,Cardoso:2003cj}, within $4$ significant digits at least, which can be used as a check for our numerical method.

\begin{figure*}
\begin{center}
\begin{tabular}{c}
\hspace{-5mm}\includegraphics[width=0.95\textwidth]{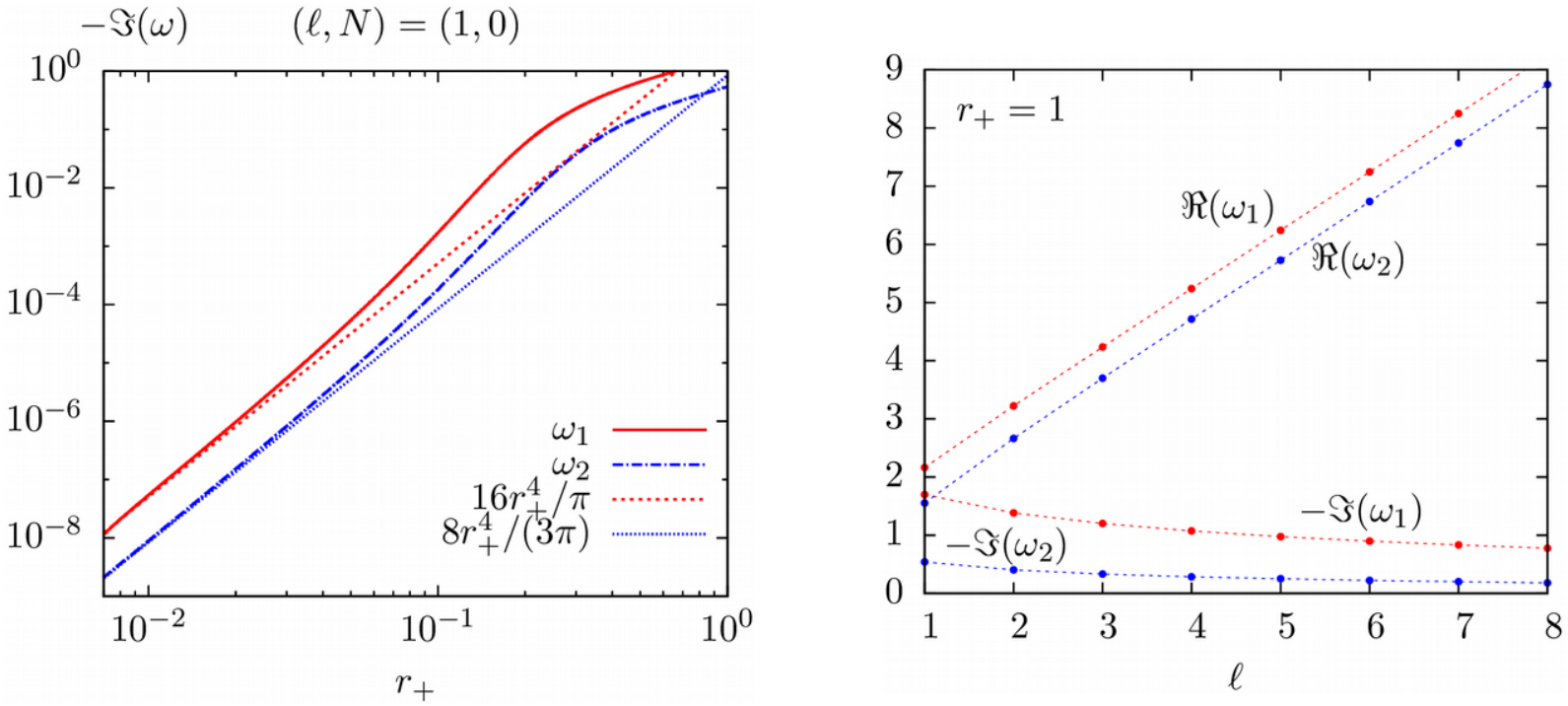}
\end{tabular}
\end{center}
\caption{\label{Mf} {\em Left:} Comparison of the imaginary part for quasinormal frequencies between the analytical approximation of small BHs (thin dashed lines) and the numerical data (thick lines) for the fundamental modes of each branch of solutions. Note that the double logarithmic scale is used in this panel. {\em Right:} Effect of the angular momentum quantum number $\ell$ on the quasinormal frequencies for intermediate size BHs with $r_+=1$, and $N=0$. The red line is for $\omega_1$ and the blue line is for $\omega_2$.}
\end{figure*}

For small BHs, it can be shown by a perturbative analytical matching method~\cite{wangherdeiro} that the real part of the frequencies approaches the normal modes of empty $AdS$~\cite{Konoplya:2002zu}, given by Eqs~\eqref{normalmode1} and~\eqref{normalmode2} for the two different BCs, while the imaginary part for both modes approaches zero as
\begin{equation}
-\omega_{j,I} \propto r_+^{2\ell+2}\;,\nonumber
\end{equation}
which also seems to be a general feature for any type of perturbation~\cite{Berti:2009wx}. In Fig.~\ref{Mf}, left panel, we display the numerical data (thick lines) for the fundamental modes of each branch against the leading behaviour obtained from the perturbative matching method~\cite{wangherdeiro} and find a good agreement for small $r_+$, which can be used as another check for our numerical method.

\begin{table}
\caption{\label{EM1} Quasinormal frequencies of the Maxwell field on Schwarzchild-$AdS$. Some fundamental modes are shown, for different BH sizes $r_+$ and for the two sets of modes.}
\begin{ruledtabular}
\begin{tabular}{ l l l }
$r_+$ & $\omega_1 (\ell=1)$ & $\omega_2 (\ell=2)$ \\
\hline\\
0.2 & 2.6384 - 5.7947$\times 10^{-2}$ i & 2.9403 - 1.0466$\times 10^{-4}$ i\\
0.5 & 2.2591 - 0.6573 i & 2.7804 - 0.07549 i\\
0.8 & 2.1758 - 1.2870 i & 2.6923 - 0.2721 i\\
1.0 & 2.1630 - 1.6991 i & 2.6647 - 0.4061 i\\
5.0 & 0 - 8.7948 i & 0 - 5.0528 i\\
10 & 0 - 15.5058 i & 0 - 13.8198 i\\
50 & 0 - 75.0958 i & 0 - 74.7533 i\\
100 & 0 - 150.048 i & 0 - 149.876 i\\
\end{tabular}
\end{ruledtabular}
\end{table}

In Table~\ref{EM2}, consider intermediate size BHs to exemplify the effect of the angular momentum quantum number $\ell$ on the frequencies; we have checked the effect of varying $\ell$ is qualitatively similar for small BHs. As one can see, for both modes, the real (imaginary) part of quasinormal frequencies increases (decreases) in magnitude as $\ell$ increases. These behaviours are more clearly shown in the right panel of Fig.~\ref{Mf}. Observe that the increasing of the real part of the frequency with $\ell$ is qualitatively similar to the one observed for empty $AdS$.

Finally, let us remark that, in the above, we have focused on fundamental modes because, on the one hand, our main interest has been to explore the new set of modes which arises even for $N=0$ and, on the other hand, since these low lying modes are expected to dominate the late time behavior of time evolutions.


\begin{table}
\caption{\label{EM2} Same as Table \ref{EM1}, but fixing now the BH size to be $r_+=1$. Some fundamental modes are shown, considering different angular momentum quantum number $\ell$.}
\begin{ruledtabular}
\begin{tabular}{ l l l }
$\ell$ & $\omega_1$ & $\omega_2$ \\
\hline\\
1 & 2.16302 - 1.69909 i & 1.55360 - 0.541785 i\\
2 & 3.22315 - 1.38415 i & 2.66469 - 0.406058 i\\
3 & 4.23555 - 1.20130 i & 3.69923 - 0.334088 i\\
4 & 5.23994 - 1.07445 i & 4.71659 - 0.286828 i\\
5 & 6.24294 - 0.97775 i & 5.72784 - 0.252025 i\\
6 & 7.24598 - 0.89976 i & 6.73632 - 0.224622 i\\
7 & 8.24941 - 0.83447 i & 7.74335 - 0.202110 i\\
8 & 9.25327 - 0.77838 i & 8.74952 - 0.183072 i\\
\end{tabular}
\end{ruledtabular}
\end{table}



\section{Discussion and Final Remarks}
\label{discussion}
In this paper we have proposed that perturbations of asymptotically $AdS$ BHs should be considered using VEF BCs. This is a simple physical principle based on the perspective that the asymptotic $AdS$ boundary acts like a perfectly reflecting mirror. We have constructed a framework for implementing perturbations of Kerr-$AdS$ BHs, with these BCs in the Teukolsky formalism and illustrated how VEF BCs can lead to new results for the specific case of a Maxwell field, even for the simpler Schwarzschild-$AdS$ background. Indeed, we have found that there are two branches of quasinormal modes, one of which has been studied in~\cite{Cardoso:2001bb,Cardoso:2003cj,Berti:2003ud}, and another which is new.

The new branch is actually isospectral to the old branch for empty $AdS$, except for the $\ell=1$ mode of $\omega_2$. This isospectrality is broken when BH effects are taken into account. To establish this we presented a numerical analysis of the quasinormal frequencies for different BH size. The breakdown of isospectrality is more pronounced in the small and intermediate BH size regimes; for large BHs two modes are again almost isospectral.

An interesting aspect is that for both small and intermediate size BHs, the imaginary part of new modes is always smaller than that of the old modes, which implies a longer decay time scale.  For small BHs, one can show analytically that the real part of both modes approach the corresponding normal modes, while the imaginary part is proportional to $r_+^{2\ell+2}$ (with different proportionality constants). Furthermore, we also studied the effect of the angular momentum quantum number $\ell$ on the frequencies. The real (imaginary) part of both modes increases (decreases) in magnitude as the angular momentum quantum number $\ell$ increases.

We would like to stress that the VEF BCs, in the case of spherically symmetric backgrounds, can be applied not only in the Teukolsky formalism, but also in the Regge-Wheeler formalism. We have checked that if one imposes VEF BCs in the latter formalism for Maxwell perturbations of Schwarzschild-$AdS$ BHs - instead of vanishing field BCs~\cite{Cardoso:2001bb} - we get the same two sets of quasinormal frequencies that we have obtained in the Teukolsky formalism.

Since the formulation we have presented can be applied to other spin fields, we have checked that, for a scalar field, the VEF BCs reduces precisely to the commonly used Dirichlet BCs. It would be interesting to apply this formulation to other spin fields, especially for the gravitational field.

Turning on the angular momentum of the BHs, the Robin BCs on the Teukolsky variables for Kerr-$AdS$ BHs can be used to study also superradiant instabilities and vector clouds of the Maxwell field. Work to investigate these aspects is underway and we hope to report on them soon~\cite{wangherdeiro}.

\bigskip

\noindent{\bf{\em Acknowledgements.}}
It is a pleasure to thank Jo\~ao Rosa for discussions and suggestions in the initial stage of this work. M.W. and M.S. are funded by FCT through the grants SFRH/BD/51648/2011 and SFRH/BPD/69971/2010. C.H. acknowledges support from the FCT IF programme. The work in this paper is also supported by the grants PTDC/FIS/116625/2010, CIDMA strategic project UID/MAT/04106/2013 and  NRHEP--295189-FP7-PEOPLE-2011-IRSES.

\bibliographystyle{h-physrev4}
\bibliography{BCs}


\end{document}